\documentclass[12pt]{article}

\usepackage{scicite}
\usepackage{times}
\usepackage{graphicx}%
\usepackage[version=4]{mhchem}
\usepackage{seqsplit}
\usepackage{commath}
\usepackage{setspace}
\usepackage[displaymath]{lineno}

\newenvironment{sciabstract}{%
\begin{quote} \bf}
{\end{quote}}

\title{Topological Singularities in Metasurface Scattering Matrices: From Nodal Lines to Exceptional Lines}

\author{Jingguang Chen,$^{1,2\dagger}$ Wenzhe Liu,$^{1,2,3\dagger\ast}$ Jiajun Wang,$^{2}$ Ruo-Yang Zhang,$^{3}$\\ Xiaohan Cui,$^{3}$ Fang Guan,$^{1,2}$  Lei Shi,$^{1,2}$ Jian Zi,$^{2\ast}$ and C. T. Chan$^{3\ast}$
\\
\normalsize{$^{1}$Institute for Nanoelectronic Devices and Quantum Computing,}\\
\normalsize{Fudan University, Yangpu District, Shanghai, 200433, China}\\
\normalsize{$^{2}$State Key Laboratory of Surface Physics, Key Laboratory of Micro- and}\\
\normalsize{Nano-Photonic Structures (Ministry of Education), and Department of Physics,}\\
\normalsize{Fudan University, Yangpu District, Shanghai, 200433, China}\\
\normalsize{$^{3}$Department of Physics, The Hong Kong University of Science and Technology,}\\
\normalsize{Clear Water Bay, Kowloon, Hong Kong, 999077, China}\\
\\
\normalsize{$^\ast$To whom correspondence should be addressed; E-mail:  wzliu@fudan.edu.cn;}\\
\normalsize{jzi@fudan.edu.cn; phchan@ust.hk.}\\
\normalsize{$^\dagger$These authors contributed equally to this work.}
}

\date{}

\begin{document}


\baselineskip24pt


\maketitle


\begin{sciabstract}
  Topological properties of photonic structures described by Hamiltonian matrices have been extensively studied in recent years. Photonic systems are often open systems, and their coupling with the environment is characterized by scattering matrices, which can exhibit topological features as well. In this work, we uncover that topological singularities can be manifested in the scattering matrices of two-dimensional periodic photonic systems with open boundaries in the third dimension, introducing a new topological approach to describe scattering. We elaborate the importance of symmetry and demonstrate that mirror symmetry gives rise to the formation of diabolic points and nodal lines in the three-dimensional frequency-momentum space, which transform into exceptional points and lines in the presence of material loss. These topological features in the eigenvalue structure of the scattering matrix manifest as vortex lines in the cross-polarization scattering phase, providing a direct link between the eigen-problem and observable scattering phenomena in the frequency-momentum space. We demonstrate these phenomena numerically and experimentally using a reflective non-local metasurface. These findings extend the concept of topological singularities to scattering matrices and pave the way for novel photonic devices and wavefront engineering techniques.
\end{sciabstract}



Topological photonics has emerged as a vibrant field, leveraging concepts from topological insulators and other condensed matter systems. In these closed condensed matter systems, the properties of modes, such as band structures, are often described using Hamiltonian matrices, which can exhibit topological singularities such as Dirac cones and nodal lines (NLs). Lifting these singularities leads to exotic transport effects like robust boundary states \cite{qi2011topological, ochiai2009photonic, lu2014topological, gao2015topological, he2016photonic, yang2017direct, parappurath2020direct, liu2020observation, yang2020observation, guo2021experimental} and topological lasing \cite{ota2018topological, gao2020dirac, shao2020high, yang2022topological}.

However, photonic systems are inherently open, allowing modes to couple with the continuum in free space. To account for this openness, one can apply an effective Hamiltonian approach derived from the Feshbach-Fano method, where the system's coupling to the continuum is treated as ``radiative loss'', that is an imaginary part of the Hamiltonian \cite{feshbach1958unified, fano1961effects, miroshnichenko2010fano}. In the context of topological photonics, when this imaginary part is neglected, topological singularities such as Dirac cones and NLs can still be observed. Including the imaginary part transforms the Hamiltonian into a non-Hermitian form, giving rise to exceptional points (EPs) which may form exceptional lines (ELs). EPs and ELs play crucial roles in fascinating physical phenomena like non-Hermitian skin effects \cite{yao2018non, shen2018topological, longhi2019probing, okuma2020topological, zhou2023observation, zhang2023electrical}, and they enable novel optical functionalities like enhanced sensing \cite{wiersig2014enhancing, chen2017exceptional, hodaei2017enhanced, hokmabadi2019non, li2023exceptional} and topological phase winding \cite{lawrence2014manifestation, park2020observation, song2021plasmonic, ding2022non}.

While the Feshbach-Fano method provides important insights, it remains an approximation. A more precise description of open photonic systems is afforded by the scattering matrix formalism, which does not rely on such approximations. The scattering matrix can be compared to a time evolution operator, $\exp (-i \mathcal{H} \chi)$, where $\mathcal{H}$ is a Hamiltonian and $\chi$ is the evolution parameter. This raises an intriguing question: can topological singularities like NLs or ELs appear in the scattering matrices of open photonic systems?

In this work, we uncover the existence of topological singularities in the scattering matrices of open periodic photonic systems like non-local metasurfaces, analogous to band structure degeneracies. We discover that the unitary scattering matrix of a system with no material loss exhibits diabolic points (DPs) protected by mirror symmetry. These DPs form continuous nodal lines (NLs) in the frequency-momentum space. Herein, unitarity in a scattering matrix is analogous to Hermiticity in a Hamiltonian matrix. Incorporating material loss breaks unitarity, leading to the splitting the NLs into pairs of ELs, which have gained strong attention in flat optics \cite{lawrence2014manifestation, park2020observation, song2021plasmonic, ding2022non, he2023scattering, baek2023non, yang2024creating}. Remarkably, these singularity points and lines give rise to topological darkness \cite{kravets2013singular, malassis2014topological, guo2021structured, ermolaev2022topological, tselikov2023topological} and phase vortex lines in the cross-polarization scattering phase maps. We experimentally demonstrate the generation of spatiotemporal vortices based on these singularities. Our findings extend the concepts of topological photonics to the scattering matrices inherent in open photonic systems, revealing new avenues for exploration and application.

\section{Basic Principles}
\begin{figure*}[htpb!]
  \centering
  \includegraphics[scale=0.9]{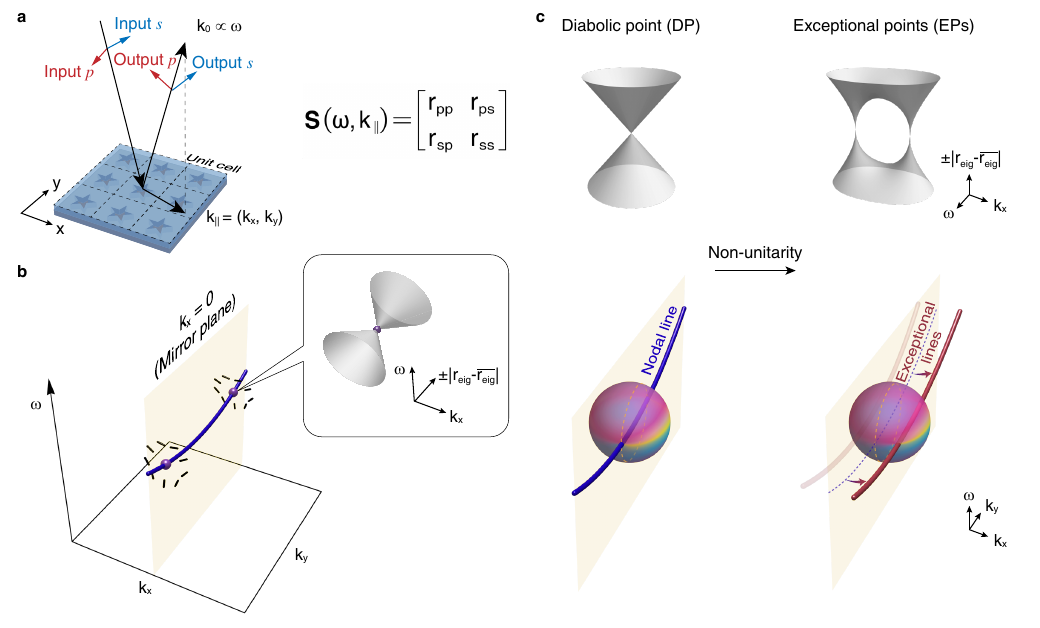}
  \caption{Degeneracies as topological singularities of scattering matrices. \textbf{a}, Illustration of the system under investigation. A periodic structure on top of a mirror substrate is used to construct the basic $2 \times 2$ scattering matrix $\mathbf{S}$. The scattering process is studied for in-plane momentum $\mathbf{k}_\parallel$ and frequency $\omega$, without considering transmission and diffraction. This simplification confines the scattering channels to two reflecting plane waves with orthogonal polarizations. The polarization basis is naturally chosen to be $p$ and $s$ polarizations. \textbf{b}, The system's mirror symmetry about the $x = 0$ plane ensures the presence of nodal lines (NLs) in the $k_x = 0$ plane of the $\omega$-$\mathbf{k}_\parallel$ parameter space. Each NL consists of diabolic points (DPs), which are characterized by the linear intersection of eigenvalues of the scattering matrix. \textbf{c}, When material loss is considered, DPs evolve into exceptional points (EPs), resulting in the division of NLs into pairs of exceptional lines (ELs). This change is also reflected in the cross-polarization scattering phase in the $\omega$-$\mathbf{k}_\parallel$ parameter space, where the phase vortices, originally located on the NL, are displaced due to the formation of ELs.}
  \label{fig:1}
\end{figure*}
To establish the theoretical foundation and reveal the existence of topological singularities in scattering matrices, we construct an optical system with the simplest $2 \times 2$ scattering matrix -- a reflection-only periodic structure, schematically illustrated in Fig. \ref{fig:1}a. The plane-wave scattering process of such a structure is characterized by a specific in-plane wavevector $\mathbf{k}_\parallel$, and there are two orthogonally polarized scattering channels. Choosing a $p$-$s$ basis of polarization, the scattering matrix $\mathbf{S}$ of the system is expressed as:
\begin{equation}\label{eqn:Smat}
  \mathbf{S(\omega,\mathbf{k_\parallel})} =
  \left[
    \begin{array}{cc}
      r_{pp}(\omega,\mathbf{k_\parallel}) & r_{ps}(\omega,\mathbf{k_\parallel}) \\
      r_{sp}(\omega,\mathbf{k_\parallel}) & r_{ss}(\omega,\mathbf{k_\parallel})
    \end{array}
  \right].
\end{equation}

Here, $\omega$ is the real angular frequency of the scattering process, which determines the free-space wavevector $\mathbf{k}_0$. The subscripts of reflection coefficients $r_{ij}$ represent the polarization channels $p$ and $s$. The $s$ channel is perpendicular to the incident plane and parallel to the sample plane, while the $p$ channel is in the incident plane and perpendicular to the input and output free-space wavevectors.

Similar to momentum-space Hamiltonian matrices, the non-diagonal entries of the scattering matrices ($r_{\mathrm{sp}}$ and $r_{\mathrm{ps}}$) are generally non-zero for a $\mathbf{k}_\parallel$ without any symmetry, which induces inter-channel scattering. Consequently, the eigenvalues of the scattering matrices cannot form a linear crossing and there will be a ``gap'' between them.

However, symmetry may make things different. Assuming that our system has mirror symmetry about $x = 0$, if $\mathbf{k}_\parallel$ is in the corresponding momentum-space mirror plane $k_x = 0$, the scattering matrix must be invariant under a reflection operation:
\begin{equation}\label{eqn:SmatSymmetry}
  \begin{gathered}
    \mathbf{S(\omega,\mathbf{k_\parallel})} = \mathbf{M} \mathbf{S(\omega,\mathbf{k_\parallel})} \mathbf{M}^{-1},\
    \mathbf{M} =
    \left[
      \begin{array}{cc}
        1 & 0 \\
        0 & -1
      \end{array}
    \right]
    \ (\text{even}),
    \ \text{or}\
    \mathbf{M} =
    \left[
      \begin{array}{cc}
        -1 & 0 \\
        0 & 1
      \end{array}
    \right]
    \ (\text{odd}),
    \\
    \Leftrightarrow
    r_{ps} = r_{sp} = 0.
  \end{gathered}
\end{equation}
Here, ``even'' and ``odd'' refer to the eigenvalues of the reflection symmetry operator with respect to the $x = 0$ plane. An ``even'' symmetry corresponds to an eigenvalue of $+1$, maintaining the input/output vector's direction, while an ``odd'' symmetry results in an eigenvalue of $-1$, reversing its direction.  This mirror symmetry enforces zero inter-channel coupling between the even-mirror-parity and odd-mirror-parity channels, resulting in a diagonal scattering matrix ($r_{ps} = r_{sp} = 0$). Consequently, the polarization channels $p$ and $s$ become the eigen-polarization channels, with $r_{pp}$ and $r_{ss}$ representing the eigenvalues of the scattering matrix.

If we further assume that there is no material loss, the scattering matrix must be unitary, meaning $\abs{r_{pp}} = \abs{r_{ss}} = 1$. Therefore, they can be expressed as exponentials of two unimodular real numbers: $r_{pp} = e^{i \phi_p}$ and $r_{ss} = e^{i \phi_s}$. For each $k_y$ with $k_x = 0$, we can tune $\omega$ to make $\phi_p$ ($r_{pp}$) cross $\phi_s$ ($r_{ss}$) linearly, leading to the existence of a diabolic point (DP) in the $\omega-\mathbf{k}_\parallel$ space. If we move away from the mirror plane, the crossing is lifted. It results in conical eigenvalue surfaces. Due to the unitarity of the matrix, the two eigenvectors of the scattering matrix remain orthogonal to each other. This is the topological degeneracy we are looking for.

Note that, $\phi_p = \phi_s$ can be satisfied on continuous curves in the $k_x = 0$ plane since we have two real number degrees of freedom, $\omega$ and $k_y$. This implies that there will be DPs forming NLs in the mirror plane, as illustrated in Fig. \ref{fig:1}b. We expect topological behavior around a NL similar to that in Hamiltonian matrices -- the eigenvectors of the scattering matrix will form polarization vortex lines around the NL \cite{burkov2011topological, lim2017pseudospin, lin2017line, yang2020observation, tiwari2020non, guo2021experimental, guo2023singular}, and such vortex lines in the eigenvectors will lead to vortex lines of cross-polarization scattering phase [Pancharatnam-Berry (PB) phase] that are robustly pinned in the mirror-symmetric $k$ plane on the NL, as shown in the left panel of Fig. \ref{fig:1}c.

Material loss is difficult to avoid in practical structures. When material loss is considered, energy is no longer conserved during the scattering process, breaking the unitarity of the scattering matrix. Consequently, the orthogonality between the eigenvectors of the scattering matrix is lost, allowing the existence of exceptional points (EPs) where both eigenvalues and eigenvectors coalesce \cite{heiss2012physics}. EPs in scattering matrices with circularly polarized eigenvectors, known as ``chiral exceptional points'', have been discussed by M. Lawrence et al. \cite{lawrence2014manifestation, park2020observation, song2021plasmonic, ding2022non, he2023scattering, baek2023non, yang2024creating}, but their origins have not been fully uncovered. In fact, EPs will emerge from DPs, such as those in the lossless case discussed above, as illustrated in Fig. \ref{fig:1}c. This is analogous to EPs in eigen-energy levels emerging from DPs in Hamiltonian matrices \cite{kirillov2005unfolding, kozii2017non, zhou2018observation, xu2023subwavelength}. Here, the NL will evolve into a pair of exceptional lines (ELs) that are mirror-symmetric about $k_x = 0$. Correspondingly, the robustness of the PB-phase vortex induced by the topological singularity is reduced. The phase vortex will shift away from the position of the original DP, and it will move around under different cross-polarization input-output setups. However, it will not vanish due to its topological nature. If we consider an input that is the same as the eigenvector of one of the EPs and the output to be orthogonal to the eigenvector, the phase vortex will coincide with the EP.

\section{Theoretical Demonstration}

\begin{figure*}[htpb!]
  \centering
  \includegraphics[scale=0.9]{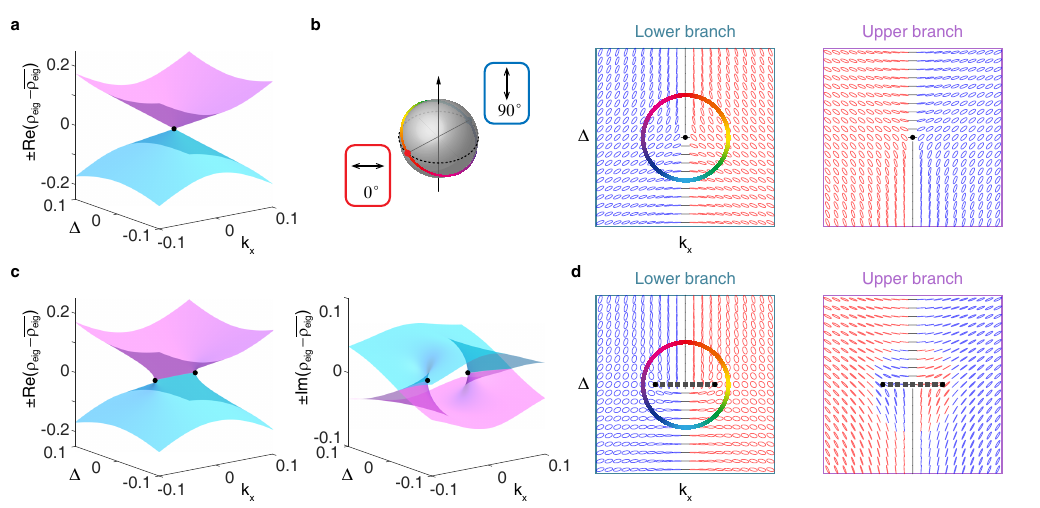}
  \caption{Demonstration of scattering-matrix topological singularities in lossless and lossy theoretical systems using small momentum ($k_x$) and frequency ($\Delta = \omega - \omega_0$) expansion. \textbf{a}, The two eigenvalue surfaces of the scattering matrix's ``core’’ with the mean value subtracted, exhibiting a DP where the two eigenvalue branches intersect linearly in a lossless system. Note that the core's eigenvalues have no imaginary part, unlike those of a true scattering matrix. \textbf{b}, Eigen-polarization maps for the two eigenvalue branches. Mirror symmetry causes a loop in the $\omega$-$k_x$ space surrounding the DP to trace a great circle on the Poincar\'e sphere. The mirror symmetry-enforced vertical and horizontal polarizations in the $\omega$-$k_x$ space map to antipodal points on the Poincar\'e sphere's equator. The mirror-symmetric polarization distribution results in an inversion symmetric trajectory on the Poincar\'e sphere that enables robust cross-polarization scattering phase vortex, provided the input/output polarization lies outside the loop. \textbf{c}, Introducing material loss splits the DP into two EPs. \textbf{d}, Correspondingly, the eigen-polarization maps reveal discontinuous changes in polarization when crossing an ``arc’’ (dashed line), arising from the self-intersecting Riemann surface structure associated with the EPs.}
  \label{fig:2}
\end{figure*}

To substantiate the above predictions and the existence of topological singularities, we apply a temporal coupled mode theory (TCMT) model. We assume the system has only two resonances in the frequency range of interest, and their eigenfrequencies are $\omega_1(k_\parallel)$ and $\omega_2(k_\parallel)$, whose subscripts mark the corresponding resonances. These resonances form photonic bands, and they can be of either even or odd mirror parity about the mirror plane $k_x = 0$. Thus, there can be three combinations of mirror parities, ``even + even'', ``odd + odd'' and ``even + odd'' (equivalent to ``odd + even''). Neglecting the material loss, we have the radiative-loss-induced decay ratios, expressed by $\gamma_{11}(k_\parallel)$ and $\gamma_{22}(k_\parallel)$, of the resonances. Note that the resonances can also couple by radiation, inducing radiation coupling ratios $\gamma_{21}(k_\parallel)$ and $\gamma_{12}(k_\parallel)$. For simplicity, we only consider resonances with $\mathbf{k_\parallel} = (k_x, 0)$, and we denote $\omega_1(0), \omega_2(0), \gamma_{11}(0), \gamma_{22}(0)$ by $\omega_1, \omega_2, \gamma_{11}, \gamma_{22}$. It is important to note that the minimum number of resonances required to demonstrate the topological singularity is two, although more resonances can be included to explore more complex scenarios which appear in practical photonic systems. This number of resonances is independent of the dimension of the scattering matrix, which is $2 \times 2$ due to the two polarizations of light. After some derivations [see Supplementary Materials], we obtain the expressions of the scattering matrix with small $\omega$-$k_\parallel$ expansions performed:
\begin{equation}\label{eqn:SmatExpSimplify1}
  \begin{aligned}
  \mathbf{S}(\omega = \omega_0 + \Delta , k_x)
  \approx
  r \left\{
  \mathbf{I}
  \ -\ \frac{1}{2}\xi
  \left[\underline{\Delta (C_1 \mathbf{I} \mp C_1 \sigma_z) + k_x (C_2 \sigma_x \mp C_3 \sigma_y)}\right]
  \right\},
  \\
  (\ - \text{ for ``even + even'' and } + \text{for ``odd + odd''})
  \end{aligned}
\end{equation}
or
\begin{equation}\label{eqn:SmatExpSimplify2}
  \begin{aligned}
  \approx
  r \left\{
  \mathbf{I}
  \ -\ \frac{1}{2}\xi
  \left[(B - i A) \mathbf{I} + \underline{\Delta (C_0 \mathbf{I} - C_1 \sigma_z) + k_x (C_2 \sigma_x - C_3 \sigma_y)}\right]
  \right\}.
  \\
  (\text{for ``even + odd''})
  \end{aligned}
\end{equation}
Here, $\omega_0$ is the central frequency around which we perform expansion, and it is determined by the properties of the two resonances at $k_x = 0$: $\omega_0 = (\omega_1 \gamma_{22} + \omega_2 \gamma_{11})/(\gamma_{22} + \gamma_{11})$ for the ``even'' + ``even'' and ``odd + odd'' parity combinations, or $\omega_0 = (\omega_1 \gamma_{22} - \omega_2 \gamma_{11})/(\gamma_{22} - \gamma_{11})$ for the ``even + odd'' case. $r$ is the direct reflection coefficient under normal incidence, while $\xi$ is a frequency-dependent factor that is not expanded since it does not affect the matrix's eigen-solutions. $A$, $B$ and $C_i$ ($i = 0,1,2,3$) are real numbers, among which $C_i$ is a number related to the $i$-th Stokes parameter of the resonances' eigen-radiation. $\sigma_{x,y,z}$ are the Pauli matrices.

One would observe that the ``core'' of the scattering matrix determining the eigen-solutions (the underlined parts) only has real and linear dependence on two degrees of freedom, the frequency $\Delta$ and wavevector $k_x$, for all the combinations. Consequently, the eigenvalues have linear dependence on $\Delta$ and $k_x$ and will intersect at $\Delta = k_x = 0$, i.e., a DP is formed. As an example, we choose the ``odd + odd'' parity combination with $C_{0,1,2,3}=1$ and plot the eigenvalues of the core, as visualized in Fig. \ref{fig:2}a. Note that the mean value of the eigenvalues has been subtracted for better visualization. A DP is clearly exhibited. We also plot the eigenvector polarization maps for the two eigenvalue branches as polarization ellipses in Fig. \ref{fig:2}b. The ellipses show topological winding behavior around the DP in the frequency-momentum space. If we project a circular loop around the singularity in the $\Delta$-$k_x$ space onto the Poincar\'e sphere, we will find two antipodal points fixed on the equator, which are the vertical and horizontal polarizations enforced by the mirror symmetry. The trajectory connecting the two fixed points will be inversion-symmetric, tracing a great circle, as a result of the mirror-symmetric polarization distribution [see the left panel of Fig. \ref{fig:2}b]. This trajectory gives rise to a robust cross-polarization scattering phase vortex pinned at the DP, provided the input/output polarization states are chosen outside this great circle of eigenvector states. Such a choice ensures a continuous phase winding rather than a sudden jump.

Based on the above lossless system, we can further take material loss into consideration by adding perturbative non-radiative decay ratios to the two resonances, $\gamma_\text{nr1}$ and $\gamma_\text{nr2}$ respectively. This will introduce extra terms to the ``core'' of the scattering matrix,
\begin{equation}\label{eqn:SmatExpSimplify3}
  \begin{aligned}
  \delta\mathbf{S}_\text{core}
  \approx
  i \left[(C_4 \mathbf{I} \mp C_5 \sigma_z) + k_x (C_6 \sigma_x \mp C_7 \sigma_y)\right].
  \\
  (\ - \text{ for ``even + even'' and } + \text{for ``odd + odd''})
  \end{aligned}
\end{equation}
As for the ``even + odd'' case, the expression shares the same form as that of the ``even + even'' case with a different definition of $C_i$. $C_{4,5,6,7}$ here are real numbers similar to $C_{0,1,2,3}$, but these numbers are proportional to $\gamma_\text{nr}$. It is obvious that the material loss introduces imaginary parts to the coefficients of the Pauli matrices, hence leading to the emergence of EPs from the original DPs. We assume $C_{4,5,6,7} = 0.05 $ and plot the eigenvalue surfaces with their eigenvector maps in Fig. \ref{fig:2}c and d. The self-intersecting Riemann surface with two degenerate points can be clearly seen, and an ``arc'' where the eigenvector ellipses abruptly change can be observed in the maps.

\section{Experimental Verification of the Singularities}

\begin{figure*}[htpb!]
  \centering
  \includegraphics[scale=0.9]{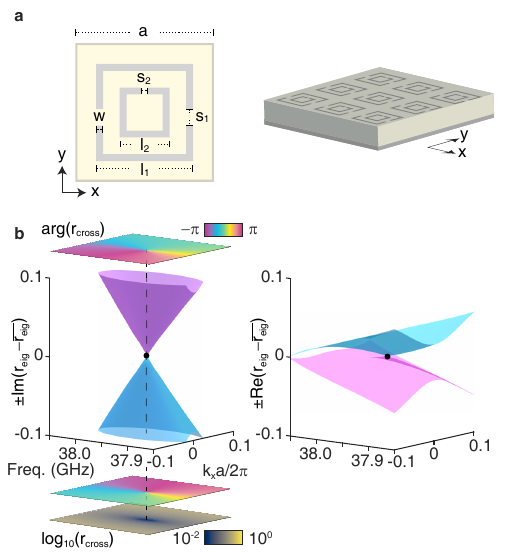}
  \caption{Non-local metasurface design for observation of topological singularities in scattering matrices. \textbf{a}, Schematic of the designed non-local metasurface. The unit cell comprises two split-ring resonators fabricated on a printed circuit board (PCB) atop a mirror substrate. The geometric parameters are: period $a = 3.810$ mm, outer ring side length $l_1 = 2.670$ mm, inner ring side length $l_2 = 1.370$ mm, linewidth $w = 0.190$ mm, outer ring split width $s_1 = 0.460$ mm, inner ring split width $s_2 = 0.190$ mm, and PCB thickness $t = 0.381$ mm. \textbf{b}, Simulated eigenvalues of the scattering matrix for the designed metasurface, neglecting the loss of the PCB. The real and imaginary parts of the eigenvalues, with the mean value subtracted, are depicted. The analysis focuses on a representative DP near 38 GHz, chosen for clarity among numerous DPs present in the scattering matrix. The map above the eigenvalue surfaces shows the cross-polarization phase for $45^\circ$ to $-45^\circ$ linear polarization conversion. The two maps below display the $-45^\circ$ to $45^\circ$ cross-polarization phase (top) and the amplitude of the cross-polarized reflection coefficient (bottom).}
  \label{fig:3}
\end{figure*}

To validate our theoretical predictions, we designed a non-local metasurface structure that operates in the microwave range, enabling the observation of topological singularities in scattering matrices. The metasurface is composed of an array of split-ring resonators fabricated on a printed circuit board (PCB) with a mirror substrate, as schematically illustrated in Fig. \ref{fig:3}a. The unit cell of the metasurface includes two split-ring resonators with specific geometric parameters: period $a = 3.810$ mm, outer ring side length $l_1 = 2.670$ mm, inner ring side length $l_2 = 1.370$ mm, linewidth $w = 0.190$ mm, outer ring split width $s_1 = 0.460$ mm, inner ring split width $s_2 = 0.190$ mm. Additionally, the outer ring splits are offset by $0.133$ mm towards the $-y$ direction. The PCB thickness is $t = 0.381$ mm.
In the absence of material loss, the eigenvalues (with their mean value subtracted) of the metasurface's scattering matrix is obtained by simulations, and we focus our attention on a representative DP near 38 GHz for clarity [Fig. \ref{fig:3}b]. The phase maps above and below the eigenvalue surfaces illustrate the cross-polarization phase for $45^\circ$ to $-45^\circ$ and $-45^\circ$ to $45^\circ$ linear polarization conversion, respectively. A phase singularity pinned at the DP's frequency and momentum can be observed in either map, corresponding to the zero point in the amplitude map of cross-polarized reflection coefficient. These $\pm45^\circ$ cross-polarization configurations provide the most distinct visualization of the phase vortices associated with the DPs. These results confirm the presence of DPs and their associated phase vortices as predicted by our theoretical model.

\begin{figure*}[htpb!]
  \centering
  \includegraphics[scale=0.9]{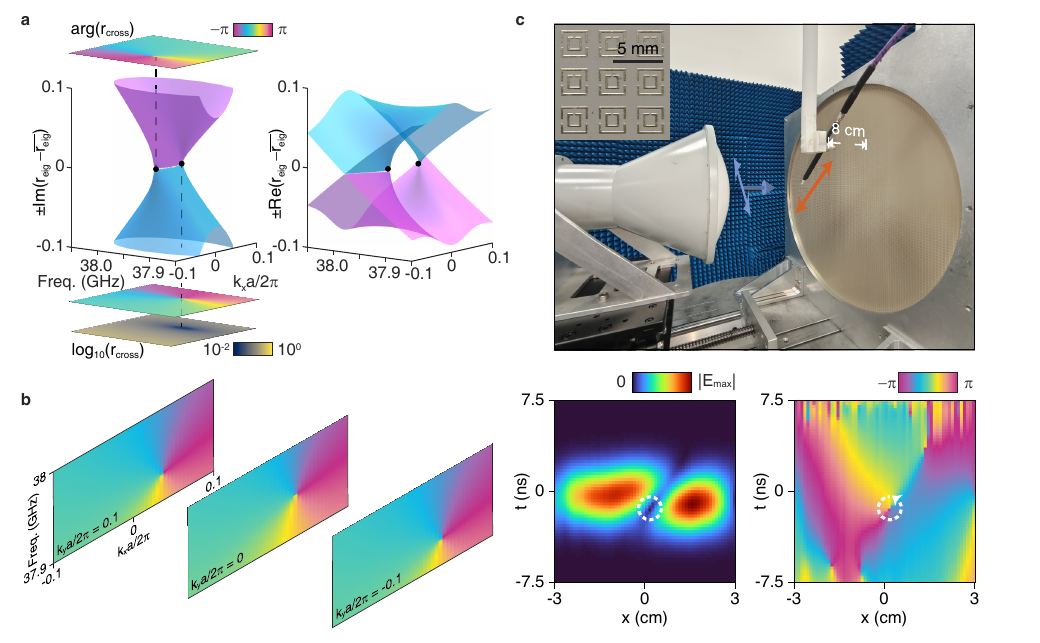}
  \caption{Observation of topological singularities via spatiotemporal vortex generation. \textbf{a}, Simulated eigenvalues of the scattering matrix for the designed metasurface, accounting for the loss in the PCB. The real and imaginary parts of the eigenvalues (with the mean value subtracted) reveal two exceptional points (EPs) and self-intersecting Riemann surfaces. The phase vortices in the phase maps shift away from $k_x = 0$ with the emergence of EPs, no longer coinciding with each other. \textbf{b}, The $-45^\circ$ to $45^\circ$ cross-polarization phase maps sliced at different values of $k_y$, marking the existence of a vortex line, corresponding to an EL, in the $\omega$-$\mathbf{k}_\parallel$ parameter space. \textbf{c}, Due to the presence of the EL in the frequency-momentum space, a spatiotemporal optical vortex can be generated and observed experimentally. A Gaussian beam with $-45^\circ$ linear polarization is impinged onto the fabricated metasurface, and the real-space frequency-domain response of the reflective sample is measured under $45^\circ$ linear polarization. After performing a frequency-domain Fourier transform, the spatiotemporal profile of the reflected pulse is obtained. The space-time intensity map of the pulse reveals a zero point, while the phase map exhibits a phase vortex, providing direct experimental evidence of the singularity line.
}
  \label{fig:4}
\end{figure*}

When we include the effects of material loss in the PCB, the simulations show a transformation in the scattering matrix's eigenvalues. Specifically, two EPs emerge by splitting from the original DP locations, accompanied by self-intersecting Riemann surfaces [Fig. \ref{fig:4}a]. This shift in the eigenvalues' imaginary and real parts demonstrates the EPs' formation due to loss. The phase vortices in the phase maps move away from $k_x = 0$, reflecting the new positions of the EPs. Fig. \ref{fig:4}b presents slices of the $-45^\circ$ to $45^\circ$ cross-polarization phase maps at different values of $k_y$. These slices reveal a vortex line in the $\omega$-$\mathbf{k}_\parallel$ parameter space, corresponding to an EL. This visualization confirms that EPs form continuous lines in the parameter space, consistent with our theoretical model.

We want to clarify that, while radiation loss in the Feshbach-Fano method introduces non-Hermiticity resulting in EPs and ELs within the Hamiltonian formalism, this type of loss is not considered true loss in the scattering matrix formalism. The scattering matrix formalism cover the entire system, including the continuum, meaning that radiation loss does not break unitarity that is analogous to Hermiticity. Therefore, only true material loss, which affects the energy conservation within the system itself, leads to the formation of EPs and ELs in the scattering matrix formalism.

To further substantiate our findings, we conducted an experimental demonstration of spatiotemporal vortex generation enabled by the singularities in the $\omega$-$\mathbf{k}_\parallel$ space. A Gaussian beam of $-45^\circ$ linear polarization was directed onto the fabricated metasurface. Then, the cross-polarization real-space frequency-domain response of the reflective sample was measured on a horizontal line 8 centimeters above the metasurface by a $-45^\circ$-polarized magnetic field probe (i.e., a $45^\circ$-polarized electric field probe). By performing a frequency-domain Fourier transform upon the measured field, we obtained the profile of the reflected cross-polarized pulse. Fig. \ref{fig:4}c shows the space-time intensity map of the pulse, revealing a zero point, while the phase map exhibits a phase vortex. This not only provides direct experimental evidence of the singularity line in the $\omega$-$\mathbf{k}_\parallel$ space that is consistent with our theoretical predictions, but also demonstrates the possible application of the topological singularities in scattering matrices.

\section{Conclusion}
In this work, we have explored topological singularities in the scattering matrices of open periodic photonic systems. We theoretically predicted and demonstrated that mirror symmetry leads to the formation of diabolic points (DPs) and nodal lines (NLs) in the frequency-momentum space. Notably, we have provided the first explanation of the origin of exceptional points (EPs) in scattering matrices, showing that the introduction of material loss transforms DPs into EPs. Our theoretical predictions were validated through the design, fabrication, and experimental characterization of a non-local metasurface structure operating in the microwave range.

Our study introduces a new paradigm in topological photonics by extending the concept of topological singularities to scattering matrices, laying the foundation for further exploration of topological phenomena in open photonic systems and offering new opportunities for designing robust photonic devices and advanced wavefront engineering techniques. Our work opens up exciting avenues for future research, such as investigating different symmetries, higher-order topological effects, or non-Abelian topological singularities in more complex systems with more scattering channels. The principles demonstrated here could also be extended to other physical systems where scattering matrices play a crucial role, e.g., acoustics and electronics.

These findings have potential applications in developing robust photonic devices for signal processing, communication, and sensing, as well as in advanced wavefront engineering. By further exploring the interplay between topology and scattering in open systems, we can uncover new functionalities and design principles for next-generation photonic devices and contribute to a broader understanding of topological phenomena in physics.



\bibliography{scibib}

\bibliographystyle{Science}



%
%

\section*{Acknowledgments}

\subsection*{Funding:}
This work is supported by National Natural Science Foundation of China (No. 12221004, No. 12234007, No. 12321161645, No. T2394480, T2394481); National Key R\&D Program of China (2022YFA1404800 and 2023YFA1406900); Science and Technology Commission of Shanghai Municipality (22142200400, 21DZ1101500, 2019SHZDZX01 and 23DZ2260100); and Hong Kong RGC (CRS\_HKUST601/23 and AoE/P-502/20).

\subsection*{Authors contributions:}
W.L. developed the theoretical framework, performed simulations, analyzed experimental data, and wrote the manuscript. J.C. conducted the experimental measurements. J.W. assisted in developing the experimental system. R.-Y.Z. and X.C. contributed to the theoretical development. F.G., L.S., J.Z., and C.T.C. supervised the work. All authors reviewed and provided input on the manuscript.

\subsection*{Competing interests:}
The authors declare no competing interests.

\subsection*{Data and materials availability}
All data are available in the main text or the supplementary materials

\section*{Supplementary materials}
Supplementary Text\\
Figs. S1-S3 \\
References

\end{document}